# Achieving Narrative Change Through AR: Displacing the "Single Story to Create Spatial Justice

Janice Tisha Samuels

The National Youth Art Movement, janice@nationalyouthartmovement.org

The ability of Augmented Reality to overcome the bias of single stories through multidimensionality is explored in the artifacts of a youth gun violence prevention project and its goal of narrative change.

## 1 INTRODUCTION

The root of social injustices and social disconnections very often come from the stories people tell about the unknown and the unfamiliar. Nigerian author Chimamanda Ngozi Adichie refers to these as the "single story." In her Ted Talk on this topic, she describes single stories as well-meaning or malicious generalizations of people and places that originate from a lack of understanding [1]. Stories that when told in repetition become biases and beliefs that inform individual and collective action or justification for inaction.

Where single stories flatten reality through the superficiality of one-dimensional or limited interpretations, Augmented Reality (AR) storytelling has the power of multidimensionality [7,8, 16]. It changes how people perceive their surroundings by adding the affordance of place-based polyvocality, the layering of multiple stories from multiple perspectives to define, develop, or resist what is, in hopes of what could be [2, 11, 12, 15]. It is this inherent quality of being able to see beyond the surface, to discover the unknown, and to transform the unfamiliar into the recognized that gives the affordance of AR the unique capability for narrative change [3, 5, 8].

For decades, the issue of gun violence in America has endured multiple single-story approaches, especially in its urban centers. For example, black people, who are disproportionately impacted by gun violence and economic inequities, have been vilified as being inherently violent, and consequently responsible for their own misfortune. However, studies have shown that white people impacted by poverty commit crimes against other white people at about the same rate as black people do against other black people [9].

Another influential single story about gun violence is that gun owners and non-gun owners have diametrically opposed attitudes toward gun safety legislation. While some news media outlets and politicians continue to stoke social divisions on the issue through repetitive messaging on Second Amendment rights, research has shown for years that most Americans are in favor of stricter gun laws [6].

These single-story narratives of gun violence that permeate media, political discourse, and influence word of mouth ignore the complexities of the issue, derailing the development and deployment of fundamental solutions to a growing public health crisis [4, 14]. Understanding that youth are the most vulnerable to the consequences of these false or superficial narratives, having little influence in mainstream media and no access to vote, the National Youth Art Movement Against Gun Violence (NYAM) formed in 2016 with a plan to use billboards and AR as a platform and place-based mechanism to empower youth-led narrative change for social and spatial justice [11].

### 1.1 NYAM Artifacts, Multidimensionality, and Spatial Justice

To mitigate the possible unconscious or conscious entrenchment of single stories about gun violence among Chicago youth and the ecosystems that surround them, NYAM developed a multi-pronged intervention that began with a call-for-art in the form of a lesson plan and ended with the city's youth leading tours of AR-enabled artwork they created and placed on billboards across their city.

The call-for-art, developed by the executive director of NYAM and local art therapists, was created to be both a tool for healing and for revealing false or superficial narratives impacting perceptions of and responses to gun violence in the city. The call-for-art instructed interested teachers on how to lead a progressive project-based lesson using art to talk about gun violence as a class; it was broken into three parts. At the close of the lesson, students could submit their finished artwork to NYAM to be considered for participation in the AR-enabled billboard art tour.

Leading from the known and familiar, parts one and two of the call-for-art asked students to 1) delve into their sense of self as it related to the known and externally unknown characteristics of their personal identity and to 2) delve into their sense of collective identity as it related to the known and externally unknown features of their neighborhoods. The goal of these activities was to examine from a place of first-hand experience and knowledge the differences between one-dimensional or limited views of people and places in comparison with the multi-dimensional. In part three of the call for art, students applied this critical thinking to the issue of gun violence, an issue familiar to some but not all. The goal in this activity was to reveal unknown biases that could prevent healing from the traumas of gun violence or developing a depth of understanding needed to enact meaningful change.

The artwork derived from the call for art expansively addressed the issue of gun violence from several vantage points: political influence, police violence, personal choices, youth death rates, and the impact on and vision of family. Via the AR billboard campaign that placed the art across numerous neighborhoods in Chicago, youth were then able to add to the collective knowledge of their city what they felt was missing in the dialogue and decision-making of residents and city leaders.

While the art prompted curiosity, investigation, and interpretation of meaning by passersby, the AR added the distinctive "voices" of black and Latino youth in the form of their interactive prevention messaging. To change the narrative of gun violence, the youth employed a spatial justice strategy of placing AR-enabled artwork focused on the impact of gun violence in neighborhoods with low levels of gun violence, which were predominantly white like Bucktown, and placing images that supported the well-being of families in neighborhoods with high levels of gun violence, which were predominantly black or Latino like Garfield Park and Humboldt Park. This strategy sought to displace single stories by exposing a range of people in the city to the unique insights of youth.

## 1.2 Conclusion

The strength of the AR-enabled billboards in the NYAM intervention was their ability to provide a platform of influence in spaces the city's youth may not have otherwise been able to reach, expanding their presence and power beyond the borders of their everyday lives and their perceived social status. AR enablement of multidimensionality provides marginalized or politically disenfranchised youth with the ability to actively disrupt biases or ignorance of the complexity in issues like gun violence by layering in the realities of other lived experiences into shared public spaces, in the hopes that fundamental solutions can become more easily recognizable and broadly supported. Children and youth advocate and criminal justice attorney Bryan Stevenson calls this aspect of social and spatial justice work becoming proximate -- getting close to the people and places that experience exclusion and social injustices to truly understand the challenges they face [13].

As access to AR tools and skills becomes more pervasive, it is the hope of NYAM that best practices and formalized methodologies for its use in developing proximity in social and spatial justice initiatives becomes established, forever disrupting the persistence and consequences of single stories.

**REFERENCES**

[1] Chimamanda Ngozi Adichie. 2009. The danger of a single story. Ted Talk. https://www.ted.com/talks/chimamanda_ngozi_adichie_the_danger_of_a_single_story?language=en
[2] Glen Cantave. How augmented reality is changing activism. 2018. Ted Talk. https://www.ted.com/talks/glenn_cantave_how_augmented_reality_is_changing_activism
[3] Brett Davison. 2022. What makes narrative change so hard? Stanford Social Innovation Review. DOI: 10.48558/xx9e-mm62
[4] Catherine Happer and Greg Philo. 2013. The role of the media in Construction of Public Belief and Social Change. Journal of Social and Political Psycjology, 1(1), 321–336, https://doi.org/10.5964/jspp.v1i1.96
[5] Alan Jenkins. 2018. Shifting the narrative: What it takes to reframe the debate for social justice in the US. The Othering and Belonging Institute at UC Berkeley. Retrieved March 1, 2024 from https://belonging.berkley.edu/shifting -narrative
[6] Jeffrey M. Jones. (2023). Majority in U.S. continues to favor stricter gun laws. Gallup. https://news.gallup.com/poll/513623/majority-continues-favor-stricter-gun-laws.aspx
[7] Xing Liu, Jieun Park, Christina Hymer, and Sherry M. B. Thatcher. 2019. Multidimensionality: A cross-disciplinary review and integration. Journal of Management, 45, 1(January 2019), 197 – 230. DOI: 10.1177/0149206318807285
[8] Hartmut Koenitz, Jonathan Barbara & Mirjam Palosaari Eladhari. 2022. Interactive digital narrative (IDN)—new ways to represent complexity and facilitate digitally empowered citizens, New Review of Hypermedia and Multimedia, 28:3-4, 76-96, DOI: 10.1080/13614568.2023.2181503
[9] Lynne C. Manzo. 2005. For better or worse: Exploring multiple dimensions of place meaning. Journal of Environmental Psychology. https://doi.org/10.1016/j.jenvp.2005.01.002
[10] Rachel E. Morgan. (2017). Race and Hispanic origin of victims and offenders, 2012-15. U.S. Department of Justice, Office of Justice Programs, Bureau of Justice Statistics. https://bjs.ojp.gov/content/pub/pdf/rhovo1215.pdf
[11] Janice Tisha Samuels, Anijo Mathew, Chantala Kommanivanh, Daniel Kwon, Liz Gomez, B'Rael Ali Thunder, Daria Velazquez, Millie Martinez, and Leah LaQueens. (2018). Art, Human Computer Interaction, and Shared Experiences: A Gun Violence Prevention Intervention. Proceedings of the Extended Abstracts of the 2018 CHI Conference on Human Factors in Computing Systems. Conference on Human Factors in Computing Systems, Montreal, Canada. Retrieved from https://dl.acm.org/citation.cfm?id=3186526.
[12] Rafael M. L. Silva, Erica Principe Cruz, Daniela K. Rosner, Dayton Kelly, Andrés Monroy-Hernández, and Fannie Liu. 2022. Understanding AR Activism: An Interview Study with Creators of Augmented Reality Experiences for Social Change. In CHI Conference on Human Factors in Computing Systems (CHI '22), April 29-May 5, 2022, New Orleans, LA, USA. ACM, New York, NY, USA 15 Pages. https://doi.org/10.1145/3491102.3517605
[13] Bryan Stevenson. Confronting Injustice. SXSW. https://www.youtube.com/watch?v=C-XIXkESlio
[14] David Peter Stroh. (2009). Leveraging Grantmaking: Understanding the Dynamics of Complex Social Systems. The Foundation Review 1, 3. https://scholarworks.gvsu.edu/cgi/viewcontent.cgi?article=1128&context=tfr
[15] Violeta Tsenova, Gavin Wood, and David Kirk. Designing with Genius Loci: An Approach to Polyvocality in Interactive Heritage Interpretation. Multimodal Technol. Interact. 2022, 6, 41. https://doi.org/10.3390/mti6060041
[16] Anya Ventura. 2021. How MIT students are transforming the art of narrative. MIT News. https://architecture.mit.edu/news/how-mit-students-are-transforming-art-narrative

# A APPENDICES

## A.1 Artifact 1: Call for Art Lesson Plan

> ### *Definitions*
>
> 1. **Identity**: the qualities, beliefs and experiences that make up an individual or community.
> 2. **Experience**: a direct observation of or participation in events.
> 3. **Perceive**: to become aware of or gain understanding through the senses, especially see, observe.
> 4. **Point of View**: a position or perspective from which something is considered; a personal way of perceiving oneself and the world.
> 5. **Joy**: a feeling of pleasure and great happiness.
> 6. **Motivation**: the reason or reasons one has for acting or behaving in a particular way.
> 7. **Fear**: an unpleasant emotion caused by the belief that someone or something is dangerous, likely to

## *Class Session 1: Introduction*

**This lesson contains reflective activities that scaffold students' understanding and evaluation of the difference between lived experience and perception and how these two different points of view can affect individual identity and the collective identity of neighborhoods. The objective of this lesson is to spark a dialogue with students about the influence and impact of stories on their everyday lives.**

> ### *Warm-Up Activity (20 minutes)*

In this activity, students will reflect on their identity. The heart in this assignment symbolizes the self. Inside and outside of the heart, students will use words to describe themselves from different points of view.

1. Provide each student with a sheet of white paper and ask them to draw a heart in the center of the paper.
2. On the outside of the heart, students will write 3 to 5 words that they feel illustrate people's perceptions of them.
3. Have students note the reasons or stories behind those perceptions in one to two sentences on the other side of their paper.
4. In the inside of the heart, students will write 3 to 5 words that they feel illustrate their perceptions of themselves.
5. Have students note the reasons or stories behind those perceptions in one to two sentences on the other side of their paper.
6. Have students circle, which qualities they wish more people knew about them.
7. Ask students to look at their work and consider how stories have amplified their motivations, fears, and joy.
8. Thinking about the power that these stories have had on their sense of self/identity or

perceptions that others have of them, have students share how they could use stories to amplify the qualities they wish more people knew about them?

## Topic Exploration (30 minutes)

In this activity, students will reflect on the unknown (inner) and known (outer) qualities of the neighborhoods they live in (e.g. people and places of support, symbols of success, acts of love and kindness, fond memories, acts of violence, causes of disability, moments of grief). To visually represent the inner and outer qualities of their neighborhood, students will choose any <u>object</u> from their neighborhood that they feel has the most significance to them. It can be an image of a particular building, a person's face (grandma, teacher, sibling, etc.), or even a symbol of a favorite season in their neighborhood like falling leaves (Autumn) or growing flowers (spring/summer).

Teacher Note: Lead into this exercise by defining and discussing the concept of identity and encouraging students to consider that their neighborhoods have an identity.

1. In groups, using presentation sized sheets of paper, students will draw an object or a set of objects that uniquely represent their neighborhood.

2. On the inside of the object(s), students will use words to show what qualities they feel are unknown to people that aren't from their neighborhood. Students will also use one or more colors to provide dimension to their object(s) by coloring them in ways that are meaningful. Students should choose colors that visually enhance the meaning of the words inside the object(s).

3. Have students note the stories behind the unknown qualities of their neighborhoods in one to two sentences on the other side of their paper.

4. On the outside of the object(s), students will use words to show qualities about their neighborhoods that are known or perceived by others. Students will again use one or more colors to provide dimension to their object(s) by coloring them in ways that are meaningful. Students should choose colors that visually enhance the meaning of the words outside the object(s).

5. Have students note the stories behind the known or perceived qualities of their neighborhoods in one to two sentences on the other side of their paper.

6. Have students circle, which qualities they wish more people knew about their neighborhoods and circle what students feel is most known or believed about their neighborhood.

7. Ask students to look at their work and consider how stories have amplified their motivations, their fears, and joy about their neighborhood? Also, ask, how are these neighborhood qualities represented in the object(s) and colors in their overall image?

8. Thinking about the influence that different colors can have on a person's feelings and reactions to a thing and the power stories can have on a person's beliefs and understanding of a place or a community, ask students to share the intention behind the colors they chose? Are the colors and shapes they chose communicating a message? If yes, who is the message for – insiders (neighbors) or outsiders (strangers)? How can stories and visuals be used to amplify the qualities they wish more people knew about their neighborhoods?

## Closing  (10 minutes)

Consider having the students write a journal response to the class activities reflecting on how they thought about stories prior to this class session and what impact this class session has had on how they evaluate the influence and impact of stories on how they see themselves, others, and the communities they live in or traverse through.

## *Class Session 2: Introduction*

**Building on the previous class session, this lesson explores what students know and don't know about gun violence, and what has informed their knowledge (e.g., news stories, social media stories, stories in books and movies, word of mouth stories, and personal experience.)**

### *Warm-Up Activity (20 minutes)*

In this activity, students will reflect on how the sources of different stories have influenced their knowledge of gun violence. The circle in this assignment symbolizes a gap of understanding or lack of knowledge.

1. In groups, using presentation sized sheets of paper, students will draw a large circle in the center of the paper
2. On the outside of the circle, students will write down what they believe they know or perceive about the causes, impacts, solutions (preventions and interventions) to gun violence.
3. Have students note the sources and stories behind their knowledge or perceptions of gun violence in one to two sentences for each of the things they listed around the circle.
4. On the inside of the circle, students will write down what they do not know, but are curious about the causes, impacts, and solutions (preventions and interventions) to gun violence.
5. Have students note the reasons for their lack of knowledge of the causes, impacts, and solutions to gun violence.
6. Have students circle what they wish they knew more about gun violence and what they feel is most known or believed about gun violence.
7. Ask students to consider and share what having a fuller story of gun violence without the gaps of understanding might achieve.

### *Art Creation (Time varies)*

In this activity, students will reflect on the deeply layered ways in which what is known and unknown about gun violence affects living in their neighborhoods or the city-at-large, and brainstorm and create a visual narrative (an image and message) that can positively influence a critical gap in understanding gun violence and its prevention and intervention through a new or fuller story. Students should intentionally consider the meaning of the objects, colors, and messages they use in their artwork and, if displayed, what places in the city these visual narratives would provide the greatest impact.

At the close of this activity, students will share their final visual narratives and explain where they would place it in their city, and the purpose they hope interactions with it will serve.

## *Closure  (10 minutes)*

To close out, have students reflect on what making their thoughts on this issue available to the public in the form of visual storytelling means to them or inspires them to consider doing with stories in the future?

*\*For the possibility of being selected to participate in an art and Augmented Reality billboard campaign for gun violence prevention, provide students interested in sharing their completed artwork with the National Youth Art Movement Against Gun Violence with additional information on the submission guidelines.*

## A.2 Artifact 2: Art Campaign Images

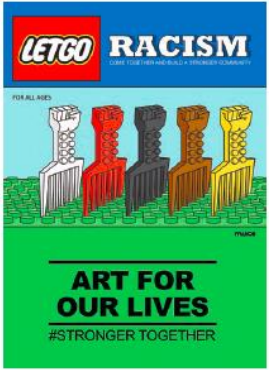
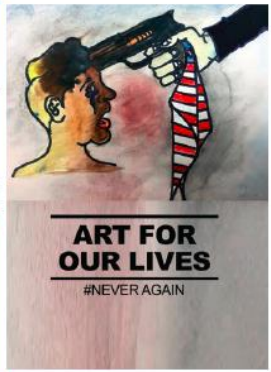
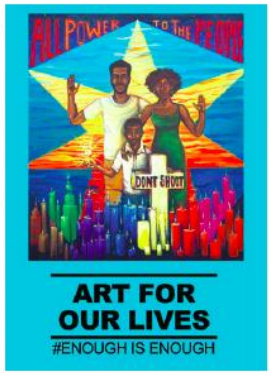
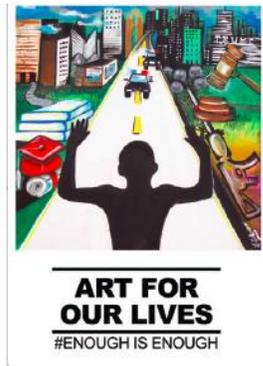
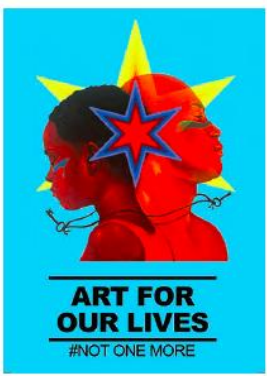
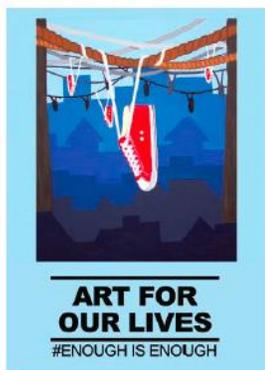
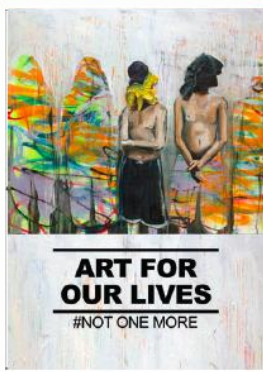
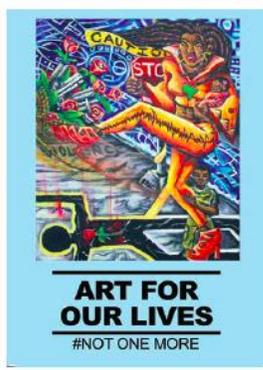